\documentclass[preprint,showkeys,secnumarabic,amsfonts,showpacs,amsmath,amssymb]{revtex4}
\usepackage[dvips]{color}
\usepackage{epsfig}
\usepackage{amsmath}
\usepackage{graphicx}

\begin{document}

\title{phase plane analysis  and statefinder diagnostic of agegraphic dark energy in 5D Brans-Dicke cosmology}

\author{Amin Salehi$^1$}\author{Hossein Farajollahi$^2$$^,$$^3$}\author{Jafar Sadeghi$^4$$^,$$^5$}\author{M.Pourali$^2$}
\affiliation{$^1$Department of Physics, Lorestan university , Lorestan, Iran}
\email{salehi.a@lu.ac.ir}
\affiliation{$^2$Department of Physics, University of Guilan, Rasht, Iran}
\affiliation{$^3$ School of Physics, University of New South Wales, Sydney, NSW, 2052, Australia}
\affiliation{$^4$ J. Sadeghi Sciences Faculty, Department of Physics, Mazandaran University,P.O. Box 47416-95447, Babolsar, Iran}
\affiliation{$^5$ Institute for Studies in Theoretical Physics and Mathematics (IPM), P.O. Box 19395-5531, Tehran,ran}

 \begin{abstract}

 We present an autonomous phase-plane describing the evolution of field equations containing an agegraphic dark energy in 5D Brans-
Dicke cosmology. To observationally verify the numerical results, we simultaneously solve the equations by constraining the model parameters with SNe Ia data. We find conditions for the existence and stability of the critical points (states of the universe) and numerically examine the cosmological parameters. We also investigate the model by means of statefinder diagnostic.

\end{abstract}


\maketitle

\section{Introduction\label{Int}}

Various cosmological observations such as Cosmic Microwave Background (CMB) \cite{Dunkley}, \cite{Komatsu}, Supernova type Ia (SNIa)\cite{Knop}, \cite{Riess1}, Weak Lensing\cite{Leauthand}, Baryon Acoustic Oscillations (BAO) \cite{Parkinson}, 2dF Galaxy Redshift Survey (2dFGRS) \cite{Cole} at low redshift and
DEEP2 redshift survey \cite{Yan} at high redshift, have provided cross-checked data to determine
cosmological parameters with high precision. These parameters imply that our approximately
$13.7$ year-old universe is nearly spatially flat, homogeneous and isotropic at large
scale, i.e. a Friedmann-Robertson-Walker (FRW) universe with zero curvature, and has entered an accelerating phase since $z \approx 0.46$ \cite{Riess1}. Moreover, according to $\Lambda CDM$ model, the universe consists of $0.046$ baryonic matter,
$0.228$ non-relativistic unknown matter, namely dark matter (DM), and a remarkable amount
of $0.726$ smoothly distributed dominant dark energy (DE) \cite{Komatsu}.
The equation of state (EoS) of DE, is the main parameter which determines
the gravitational effect of DE on the evolution of the universe, and can be measured from
observations without need to have a definite model of DE. Strong evidences imply that the
EoS of DE lies in a narrow range around $w \approx -1$ and has a smooth evolution \cite{Riess}, \cite{Amanullah}. Theoretically, one can classify the EoS of
DE with respect to the barrier $w = -1$, namely the phantom divide line (PDL )\cite{Cai}. That is,
DE with the EoS of $w = -1$ employs the cosmological constant, $\Lambda$, with a constant energy
density. The case with dynamical EoS where $w\geq -1$, is referred to as quintessence  \cite{Gonzalez}. and $w \leq -1$ corresponds to phantom energy  \cite{Carroll}, \cite{Caldwell}.

On the other hand, the problem of DE, its energy density and EoS parameter is still an unsolved problem in classical gravity and may be in the context of quantum gravity we achieve a more inclusive insight to its properties \cite{Cohen}. The holographic dark energy (HDE) model is an attempt to apply the nature of DE within the framework of quantum gravity \cite{Hsu}, \cite{Li}. The holographic principle states that the number of degrees of freedom describing the physics inside a volume (including gravity) is bounded by the area of the boundary which encloses this volume and thus related to entropy scales with the enclosing area of the system \cite{Hooft}. Since the entropy scales like the area rather than the volume, the fundamental degrees of freedom describing the system are characterized by an effective quantum field theory in a box of size L with one fewer space dimensions and with planck-scale UV cut-off $\Lambda$ \cite{Hooft}.

Among all the cosmological models, the scalar-tensor theories have been widely used to explain the late time acceleration of the universe and its relation to the HDE \cite{Cai1}--\cite{farajollahi}.

One example is Brans-Dicke (BD) theory \cite{Brans}, where the gravitational constant becomes time dependent varying as inverse of a time dependent scalar field which couples to gravity with a coupling parameter $\omega$. Many
of the cosmological problems can be successfully explained by using this theory. Alternatively, the higher dimensional theories may explain cosmic acceleration and phantom crossing \cite{mtheory7}, \cite{mtheory}. The existence of extra dimensions is required in various theories beyond the standard model of particle physics, especially in theories unifying gravity with the other fundamental forces, such as superstring and M theories \cite{mtheory3}--\cite{mtheory6}.

While the successful HDE model explains the observational data and has been studied
widely by the authors ( see for example \cite{Q}-\cite{X}), more recently, a new dark energy model, dubbed "agegraphic
dark energy" (ADE) model, has been proposed by Cai \cite{R}. The ADE is also related to the holographic
principle of quantum gravity and takes into account the uncertainty
relation of quantum mechanics together with the gravitational effect in general relativity.

Here, in  an attempt to integrate both the scalar tensor and higher dimensional theories, the 5-dim Brans-Dicke (BD) cosmology is studied with the aim to understand the agegraphic nature of dark energy in the model. We perform stability analysis and investigate the attractor solutions of the model by utilizing the 2-dimensional phase space of the theory. Also, we simultaneously best fit the model parameters with the observational data using $\chi^2$ method. This enables us to find the best fitted model parameters for the analysis of the critical points and also verification of the model with the experiment. In addition, we use the well-known statefinder parameters \cite{Sahni} the differentiate among cosmological models. These parameters are used to explore a series of DE cosmological models \cite{Alam}-\cite{H. Farajollahi}


\section{The model \label{FE}}

The 5-D Brans-Dicke action is given by,
\begin{eqnarray}
 S=\int{
d^{5}x\sqrt{^{(5)}g}\left(\frac{\phi ^2}{8\omega}
{^{(5)}R}-\frac{1}{2}g^{AB}\nabla_{A}\phi \nabla_{B}\phi
+L_M \right)},\label{act1}
\end{eqnarray}
where $\omega$ is a dimensionless coupling constant which determines the coupling between gravity
and BD scaler field, ${R}$ is 5D Ricci scaler, $\phi(x^A)$ is the BD scalar
 field and $L_{M}$ is the Lagrangian of matter field.  We assume the metric in 5-D space-time in the form of,
\begin{eqnarray}
 ds^2=-dt^2+a^2(t,y)\left(\frac{dr^2}{1-kr^2}+r^2d\Omega^2\right)+b^2(t,y)dy^2,\label{metric}
 \end{eqnarray}
where the 4-D space time is assumed to be homogeneous and isotropic (FRW  universe). In the metric, $k$ is the curvature parameter with $k = -1, 0, 1$ corresponding to open, flat, and
closed universes, respectively. The scaler field $\phi$ and the scale factors $a$ and $b$, in general are functions of $t$ and $y$.
For simplicity and plausibility, we assume the extra dimension is cyclic, i. e. the hypersurface-orthogonal space-like is a killing vector field in the underlying 5D space-time \cite{Qiang}. Hence, all fields
 are function of the cosmic time only. Note that, the functionality of scale factor $b$ on $y$, either can be eliminated by transforming to a new extra coordinate if $b$ is a separable function, and or makes no change in the following equations if $b$ is the only field that depends on $y$. Beside, in the compacting extra dimension scenarios, all field are Fourier-expanding around $y$, and henceforth one can have terms independent of $y$ to be observable, i. e. physics would thus be effectively independent of compactified fifth dimension \cite{Over}.

 Variation of the action (\ref{act1}) with respect to metric (\ref{metric}) yields the following field equation,
\begin{eqnarray}
&&\frac{3}{4\omega}\phi^2\Big(H^2+\frac{\dot{b}}{bH} H^{2}+\frac{k}{a^2}\Big)-\frac{1}{2}\dot{\phi}^2+\frac{3}{2\omega}H\dot{\phi\phi}
+\frac{1}{2\omega}\frac{\dot{b}}{bH} H\dot{\phi}\phi=\frac{\rho}{b}+\rho_B,\label{BD1}\\
&&\frac{-1}{4\omega}\phi^2\Big(2\frac{\ddot{a}}{a}+H^2+2\frac{\dot{b}}{bH} H^{2}+\frac{\ddot{b}}{b}+
\frac{k}{a^2}\Big)-\frac{1}{2\omega}(2H+\frac{\dot{b}}{bH} H)\dot{\phi\phi}-
\frac{1}{2\omega}\ddot{\phi}\phi-\frac{1}{2}\Big(\frac{1}{2}\nonumber \\
&+& \frac{1}{\omega}\Big)\dot{\phi}^2=p_B,\label{BD2}\\
&&-\frac{3}{4\omega}\phi^2\Big(2H^2+\dot{H}+\frac{k}{a^2}\Big)-\frac{1}{4\omega}\phi\ddot{\phi}-\frac{1}{4}\dot{\phi}^{2}\Big(1+\frac{1}{\omega}\Big)-\frac{3H}{4\omega}\phi\dot{\phi}=0,\\
&&\ddot{\phi}+3H\dot{\phi}+\frac{\dot{b}}{bH} H\dot{\phi}-\frac{3}{2\omega}\Big(\frac{\ddot{a}}{a}+
H^2+\frac{k}{a^2}+\beta H^{2}+\frac{\ddot{b}}{3b}\Big)\phi=0,\label{BD3}
\end{eqnarray}
where $H=\frac{\dot{a}}{a}$.
In the orthonormal basis $e^0 = dt$,$ e^i = adx^i$ and $ e^5 = bdy$, the stress-energy tensor can be
considered as.
\begin{eqnarray}
T^{A}_{B}=T^{A}_{B}|_{bulk}+T^{A}_{B}|_{brane}
\end{eqnarray}
where $T^{A}_{B}|_{bulk}$ is the energy momentum tensor of the bulk matter and
\begin{eqnarray}
T^{A}_{B}|_{bulk}=diag(-{\rho}_B,p_{B},p_{B},p_{B},q_{B})
\end{eqnarray}
The second term corresponds to the matter content in the brane (y=0)
\begin{eqnarray}
T^{A}_{B}|_{brane}=diag(-{\rho},p,p,p,0)
\end{eqnarray}
Assume that the 05 component of the energy-momentum tensor vanishes, which means that there is no
flow of matter along the fifth dimension. Therefore the nonzero elements of the 5D stress-energy tensor
are
\begin{eqnarray}
T_{00}={\rho}_B+\frac{\delta(y)}{b}\rho.
\end{eqnarray}
\begin{eqnarray}
T_{ii}=a^2{\rho}_B+a^2\frac{\delta(y)}{b}p.
\end{eqnarray}
\begin{eqnarray}
T_{55}=b^2q_B.
\end{eqnarray}
We assume that the BD scalar field $\phi$, is in power law of the scale factor $a(t)$, in the form of  $\phi\propto a^\alpha$. In the next section, we apply the ADE model in 5-D Brans-Dicke theory.
The conservation equations for the dark energy and matter field in the universe are respectively,
\begin{eqnarray}
&&\dot{\rho}_B+3H({\rho}_B+p_{B})+{\rho}_B(\frac{\dot{b}}{b})=0,\label{consq}\\
&&\dot{\rho}_m+3H\rho_m=0, \label{consm}
\end{eqnarray}

The agegraphic dark energy model with the dark energy density is given by
\begin{eqnarray}
\rho_{\Lambda}=\frac{3{n}^2M_P^2}{T^2},\label{NADE}
\end{eqnarray}
where $T$ is chosen to be the age of our universe and given by
\begin{eqnarray}
T=\int{\rm d}t=\int_0^a\frac{{\rm
d}a}{Ha}.\label{eta}
\end{eqnarray}
In the framework of Brans-Dicke cosmology, we write the agegraphic energy density of the quantum fluctuations in the
universe as
\begin{eqnarray}\label{rho1n}
\rho_{\Lambda}= \frac{3n^2\phi^2 }{4\omega T^2}.
\end{eqnarray}

\section{STABILITY ANALYSIS-PERTURBATION AND PHASE SPACE}
The structure of the dynamical system can be studied via phase plane analysis, by introducing
the following dimensionless variables,
\begin{eqnarray}\label{HL}
\Omega_m=\frac{4\omega\rho_m}{3b\phi^2
H^2}, \ \ \ \Omega_\Lambda=\frac{n^2}{H^2T^2}, \ \ \ \Omega_k=\frac{k}{H^2 a^2}, \ \ \ \
\Omega_b=\frac{\dot{b}}{bH},
\end{eqnarray}
The field equations (\ref{BD1})-(\ref{BD3}) in terms of the new dynamical variables become,
\begin{eqnarray}
\Omega_m'&=&-\Omega_m(3+2\alpha+\Omega_b+2\frac{\dot{H}}{H^2})\label{fs1}\\
\Omega_\Lambda'&=&-2\Omega_\Lambda(\frac{\sqrt{\Omega_\Lambda}}{n} +\frac{\dot{H}}{H^2})\label{fs2}\\
\Omega_k'&=&-2\Omega_k(1+\frac{\dot{H}}{H^2})\label{}\\
\Omega_b'&=&\frac{\ddot{b}}{bH^2}-\Omega_b(\Omega_b+\frac{\dot{H}}{H^2})\label{fs3}
 \end{eqnarray}
where prime " $\prime$ " means derivative with respect to $ln(a)$. In the above equations we have
\begin{eqnarray}\label{hd}
\frac{\dot{H}}{H^2}=-\frac{1}{1+\frac{\alpha}{3}}\Big(1+\frac{\alpha^2(2+\omega)}{3}+\alpha+\Omega_k\Big)
 \end{eqnarray}
  By using the Friedman constraint in terms of the new
dynamical variables:
\begin{eqnarray}
\Omega_k-\Omega_m-\Omega_\Lambda+2\alpha(1-\frac{\alpha\omega}{3})+\Omega_{b}(1+\frac{2\alpha}{3})=1
\end{eqnarray}
the Eq. (\ref{fs1})-(\ref{fs3}) reduce to,
\begin{eqnarray}
\Omega_m'&=&-\Omega_m\Big[3+2\alpha+\Omega_{b}-\frac{2}{{1+\frac{\alpha}{3}}}\Big(1+\frac{\alpha^2(2+\omega)}{3}+\alpha+\Omega_m+\Omega_\Lambda
-2\alpha(1-\frac{\alpha\omega}{3})\nonumber\\
&-&\Omega_b(1+\frac{2\alpha}{3}\Big)\Big]\label{fs4}\\
\Omega_\Lambda'&=&-2\Omega_\Lambda\Big[\frac{\sqrt{\Omega_\Lambda}}{n}-\frac{1}{1+\frac{\alpha}{3}}
\Big(1+\frac{\alpha^2(2+\omega)}{3}+\alpha+\Omega_m+\Omega_\Lambda
-2\alpha(1-\frac{\alpha\omega}{3})\nonumber\\&-&\Omega_b(1+\frac{2\alpha}{3})\Big)\Big]
\Big]\\
\Omega_b'&=&3(-1-\frac{2\alpha}{3}+\frac{2\sqrt{\Omega_\Lambda}}{n}-\Omega_b)\Omega_\Lambda+2-\frac{2\alpha+3}{1+\frac{\alpha}{3}}\Big(\nonumber\\
&-&\frac{1}{1+\frac{\alpha}{3}}\Big(1+\frac{\alpha^2(2+\omega)}{3}+\alpha+\Omega_m+\Omega_\Lambda
-2\alpha(1-\frac{\alpha\omega}{3})-\Omega_b(1+\frac{2\alpha}{3})\Big)\Big)\nonumber\\
&+&2\Omega_b(1+\frac{2\alpha}{3})+2(2+\Omega_b)\alpha+(4+\omega)\alpha^2-\Omega_b\Big(\Omega_b\nonumber\\
&-&\frac{1}{1+\frac{\alpha}{3}}\Big(1+\frac{\alpha^2(2+\omega)}{3}+\alpha+\Omega_m+\Omega_\Lambda
-2\alpha(1-\frac{\alpha\omega}{3})-\Omega_b(1+\frac{2\alpha}{3})\Big)\Big)\label{fs5}
\end{eqnarray}
It is more convenient to investigate the properties of the dynamical equations (\ref{fs4})-(\ref{fs5})
than equations Eq. (\ref{fs1})-(\ref{fs3}). In stability analysis, the above equations can be solved to find fixed points (critical points). These points are always exact constant solutions in the
context of autonomous dynamical systems. The critical points in this model are highly nonlinear and depend on the stability parameters. In addition, in stability analysis the expressions for the critical points and eigenvalues are long and cumbersome such that the usual procedure for stability analysis is not possible. To overcome the problem, we solve the above equations by best fitting the stability and model parameters and initial conditions with the observational data for distance modulus using the $\chi^2$ method. This helps us to find the solutions for the above equations and conditions for the stability of the critical points that are physically more meaningful and observationally more favored. With the simultaneous stability analysis and best fitting, we find two critical points. The best fitted stability and model parameters and initial conditions are $\alpha=0.6$, $\omega=-2.9$,$n=1.2$, $\Omega_\Lambda(0)=.7$, $\Omega_b(0)=.6$ and $\Omega_m(0)=.27$. The properties of the two best fitted critical points are given in Table I.

\begin{table}[ht]
\caption{Best fitted critical points} 
\centering 
\begin{tabular}{c c c } 
\hline 
\hline 
CP  &  $(\Omega_\Lambda,\Omega_m,\Omega_b)$ \ & $stability$ \\ [3ex] 
\hline 
\hline 
$CP1$ & \ $(0, 0, 3.5)$ &$Stable$  \\
\hline 
$CP2$ & \ $(0, 4.7, 0.9)$ \ & $Unstable $   \\
\hline 
$CP3$ & \ $(0, 0, -0.6)$ \ & $Saddle Point$   \\
\hline 
$CP4$ & \ $(0, 16.3, 6.7)$ \ & $Saddle Point$   \\
\hline 
$CP5$ & \ $(0.2, 0.2, -0.5)$ \ & $Saddle Point$   \\
\hline 
$CP6$ & \ $(0.9+2I, 0.2, -0.5+0.9I)$ \ & $Saddle Point$   \\
\hline 
$CP7$ & \ $(0.9-2I, 0.2, -0.5-0.9I)$ \ & $Saddle Point$   \\
\hline 
$CP8$ & \ $(3.6, -2.8, -1)$ \ & $Unstable$   \\
\hline 
$CP9$ & \ $(4.8, -3, -0.5)$ &$Unstable$  \\
\hline 
\end{tabular}
\label{table:1} 
\end{table}\

In Fig. 1, the trajectories leaving the unstable critical point CP1 in the past in the phase plane is shown going towards the stable critical point CP2 in the future. The best fitted model parameter trajectories is also shown by green dashed trajectory.\\

\begin{tabular*}{.5 cm}{cc}
\includegraphics[scale=.5]{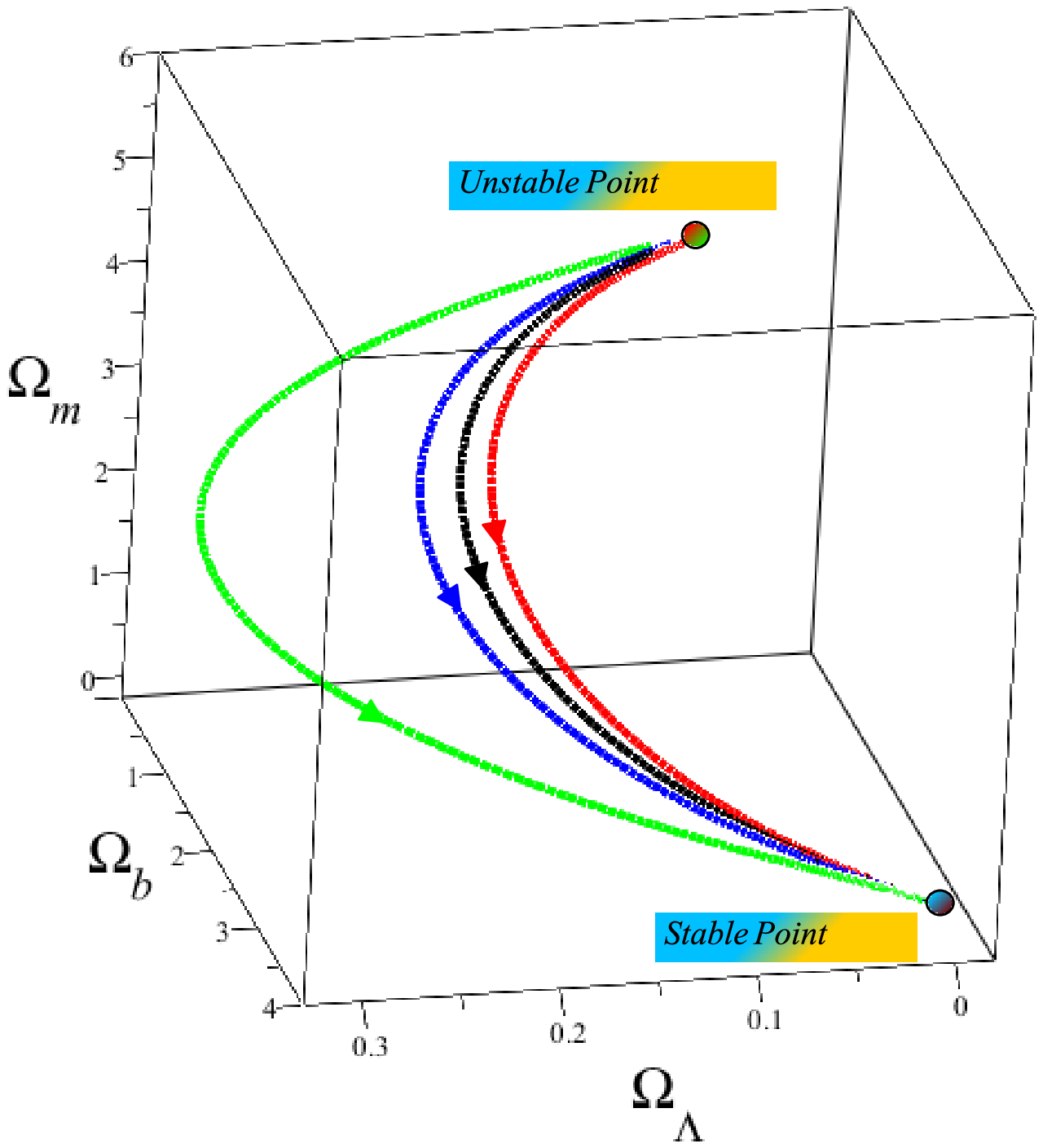}\hspace{0.1 cm}\\
\end{tabular*}\\
\begin{tabular*}{.5 cm}{cc}
\includegraphics[scale=.27]{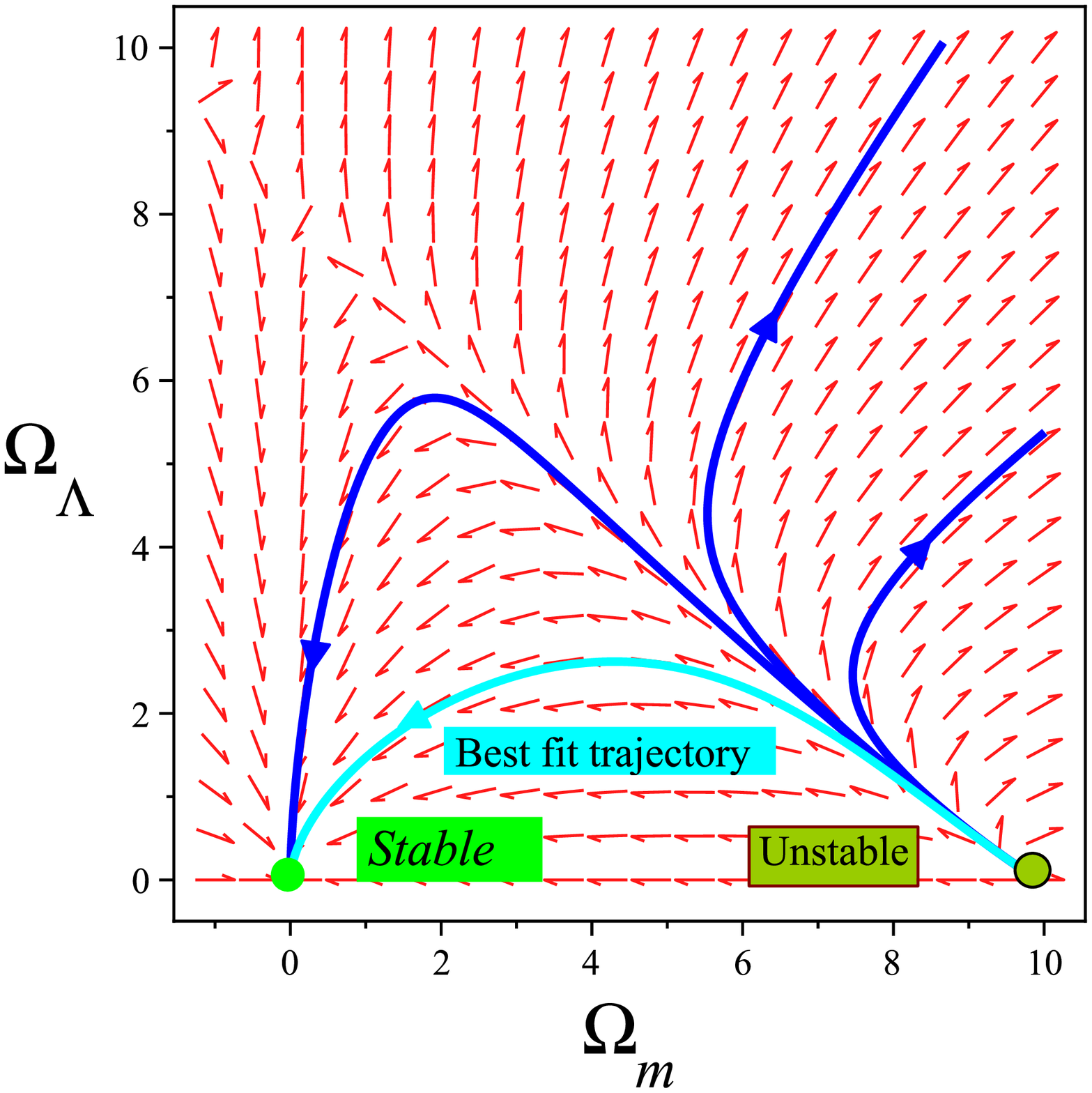}\hspace{0.1 cm}\includegraphics[scale=.27]{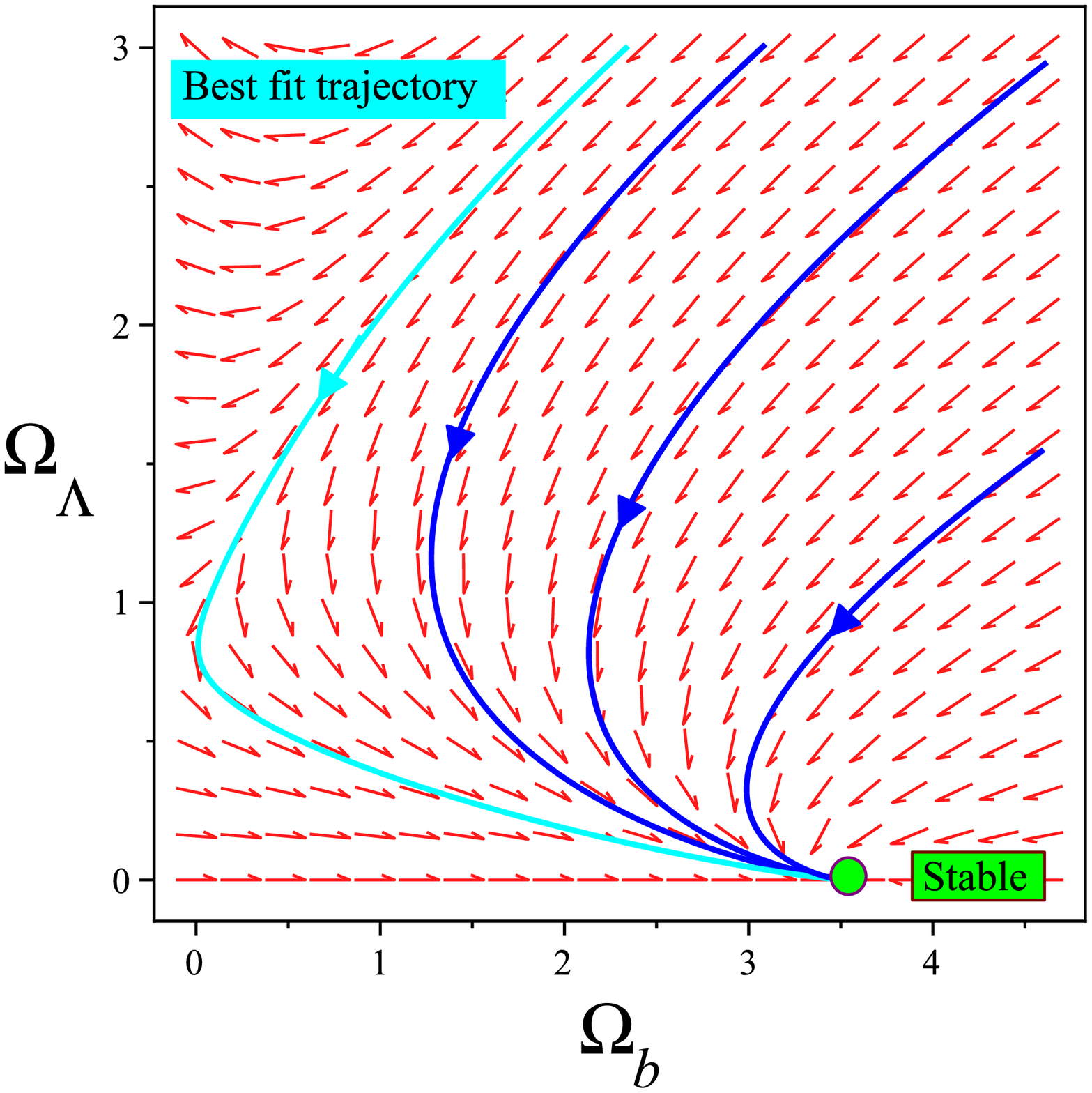}\hspace{0.1 cm}\includegraphics[scale=.27]{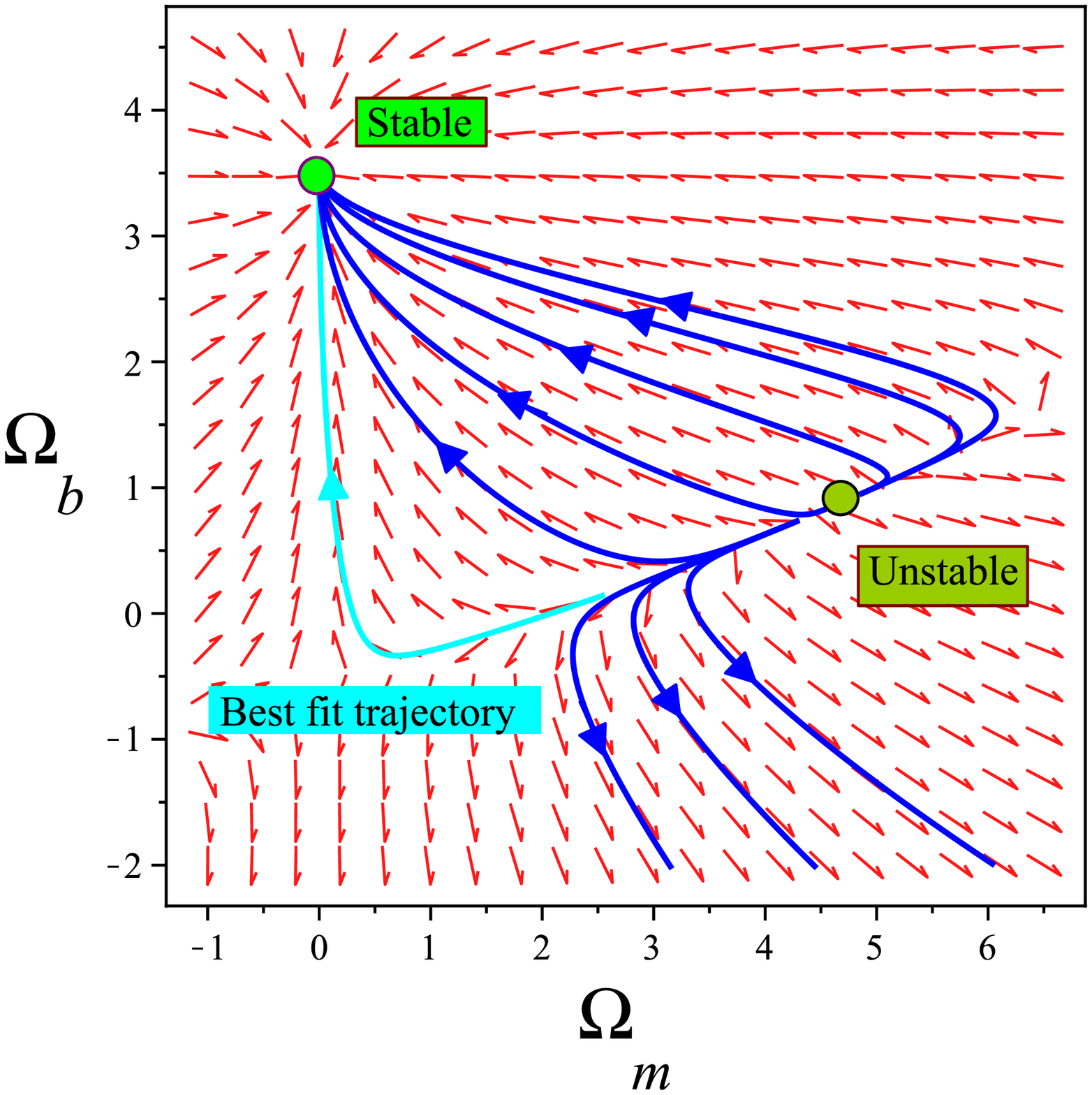}\hspace{0.1 cm}\\
Fig. 1:The 3-dim and 2-dim phase plane corresponding to the critical point\\
\end{tabular*}\\

\section{Cosmological tests and statefinders}

In order to understand the behavior of the universe and its dynamics we need to study the
cosmological parameters. We have best fitted our model with the current observational data
by the distance modulus test. The cosmological parameters analytically and/or numerically
have been investigated by many authors for variety of cosmological models. Simultaneously
best fitting the model with the observational data gives us a better understanding of the
solutions and the dynamics of these parameters.

We begin with the agegraphic energy density given by equation (\ref{rho1n}). Taking derivative of (\ref{rho1n}) with respect to cosmic time and substituting into conservation equation (\ref{consq}), in terms of new dynamical variables we obtain
\begin{eqnarray}\label{rho1n}
w_\Lambda= -1-\frac{2\alpha}{3}+\frac{2\sqrt{\Omega_\Lambda}}{3n\alpha}-\Omega_b.
\end{eqnarray}
With the best fitted stability and model parameters, the EoS parameter for agegraphic dark energy can be obtained. One has to know that the contribution of extra dimension into the formalism is via the presence of parameter $\alpha$ in the field equation for $\Omega_\Lambda$.\\

\begin{tabular*}{2.5 cm}{cc}
\includegraphics[scale=.5]{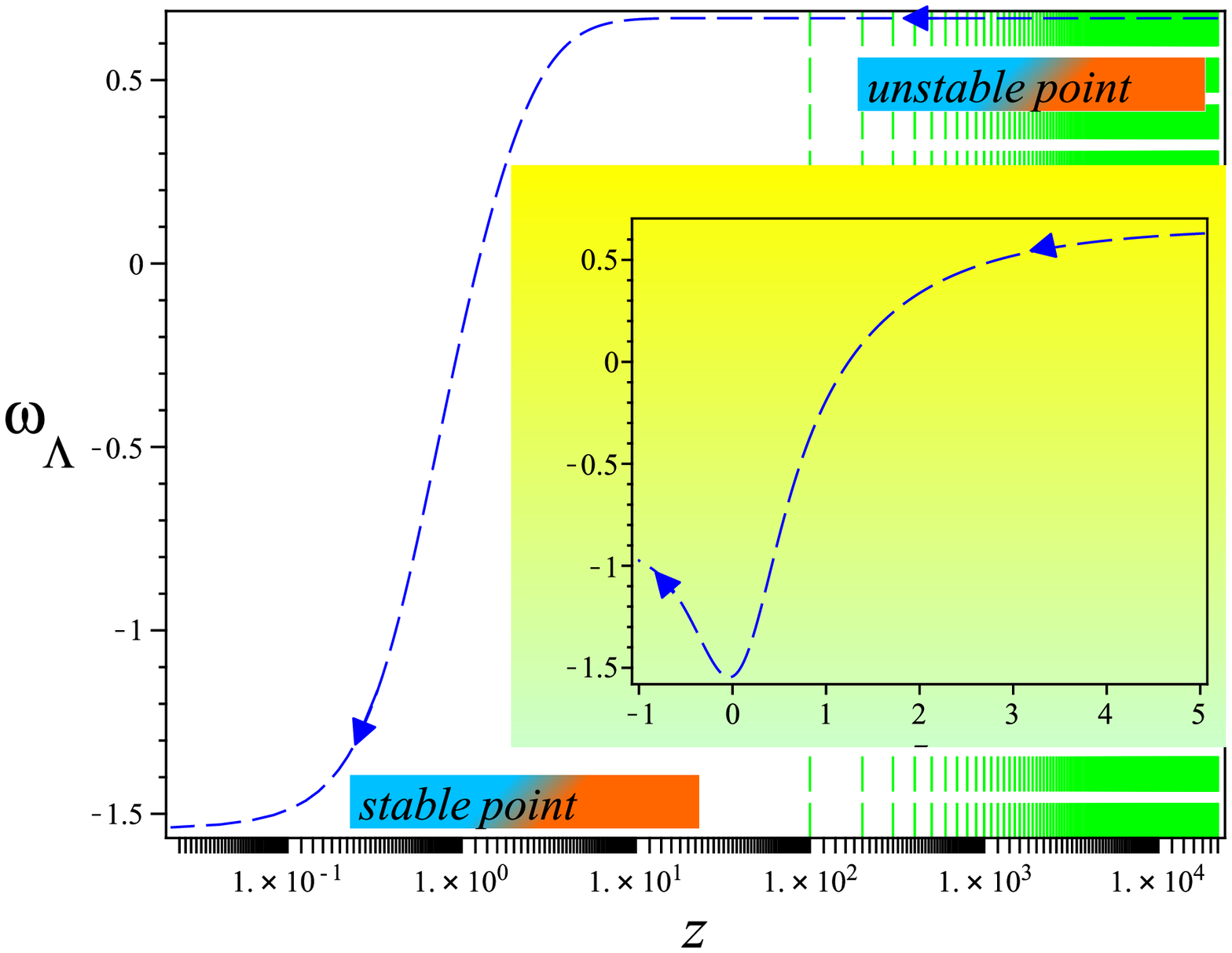}\hspace{0.1 cm}\\
Fig. 2: The best fitted EoS parameter for agegraphic dark energy, $w_{\Lambda}$, plotted as functions of redshift\\
\end{tabular*}\\

Among cosmological parameters, the effective EoS parameter and statefinders are given by
\begin{eqnarray}
w_{eff} =-1-\frac{2}{{3(1+\frac{\alpha}{3})}}\Big[(1+\frac{\alpha^2(2+\omega)}{3}+\alpha+\Omega_m+\Omega_\Lambda
-2\alpha(1-\frac{\alpha\omega}{3})
-\Omega_b(1+\frac{2\alpha}{3})\Big]
\end{eqnarray}
, $r = \frac{\ddot{H}}{H^3}-3q-2$
and $s =\frac{(r-1)}{3(q-\frac{1}{2})}$ are discussed here, where $q$ in $r$ and $s$ is the deceleration parameter
and $\frac{\ddot{H}}{H^2}$ in $r$ in terms of new dynamical variables for exponential and power law cases can
be obtained by taking derivative of $\dot{H}$.\\

In Table II, the best fitted values of the effective EoS and statefinder parameters of the two critical points are given.\\
\begin{table}[ht]
\caption{Best fitted values of effective EoS and statefinder parameters} 
\centering 
\begin{tabular}{||c |c| c| c| c| c| c| c| c| c|} 
\hline\hline 
parameter  &  $w_{eff}$ \ & $q$ &  $r$ \ & $s$ & $\omega_{\Lambda}$\\ [4ex] 
\hline 
Current Value & \ $-.85$ & $-.8$ & \ $-3.2$ & $1.065$& $-1.52$ \\
\hline 
CP1 & \ $-.74$ & $-.61$ & \ $-2.4$ & $1.065$& $-.97$ \\
\hline 
CP2 & \ $0.46$ & $1.75$ & \ $3.7$ & $1.415$& $.67$ \\
\hline 
\end{tabular}
\label{table:1} 
\end{table}\\

Fig. 3 shows the dynamics of the best fitted effective EoS parameter against redshift $z$. From graph or Table II, the best fitted current value of EoS $w_{eff}$ is $-.85$. The graph also shows that universe undergoes phantom crossing twice in future. From a stability point of view it shows that the universe starts from an unstable state in the higher redshift in the far past and tends to a stable big rip state when the the scale factor of the universe becomes infinite at a finite time in the future.\\

\begin{tabular*}{2.5 cm}{cc}
\includegraphics[scale=.5]{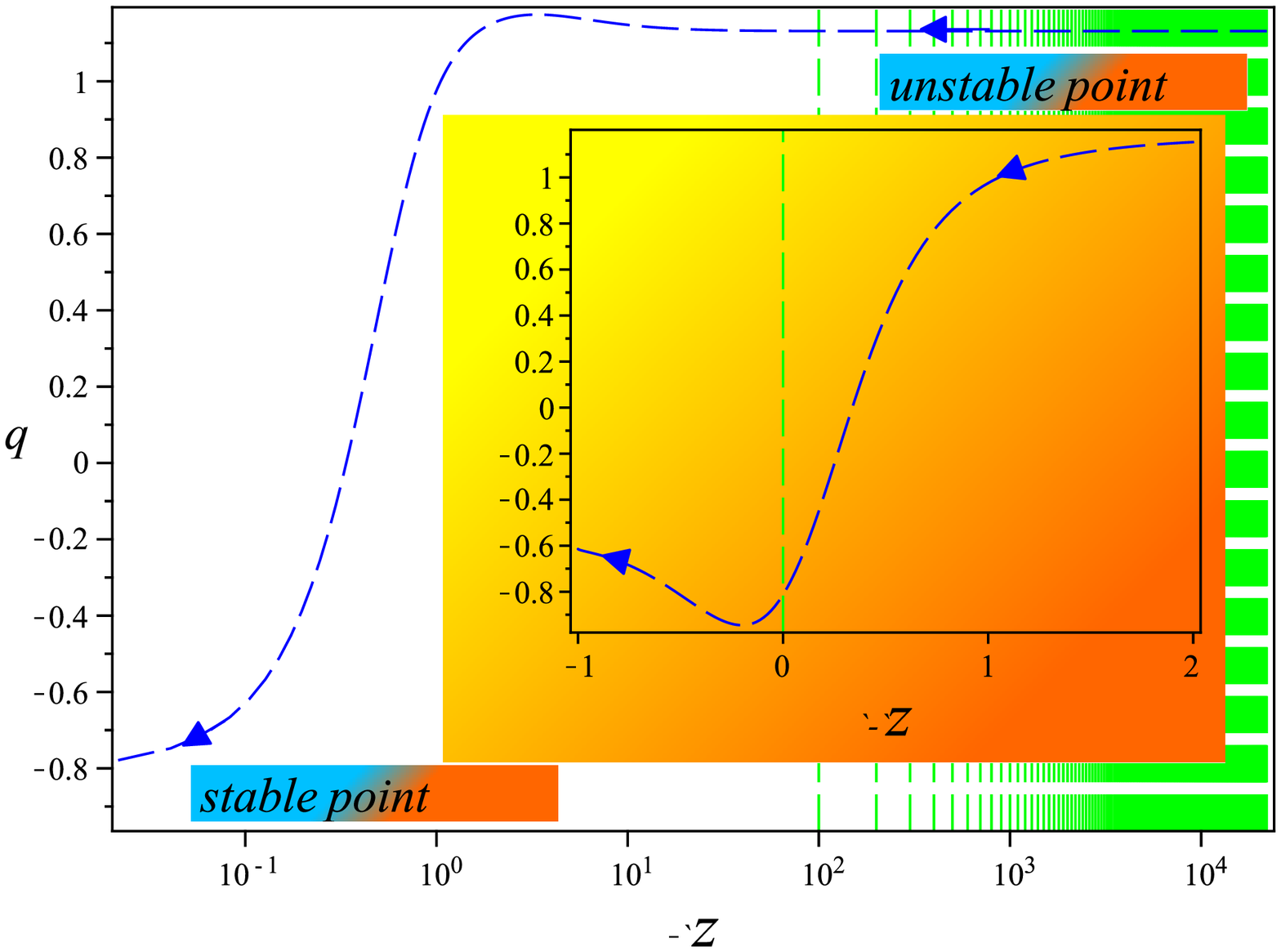}\hspace{0.1 cm}\\
Fig. 3: The best fitted effective EoS parameter $w_{eff}$, plotted as\\ functions of redshift,in the solid line) presence and dotted line)absence of the extra dimension.\\
\end{tabular*}\\

Fig 4 shows the best-fitted trajectories of the statefinder diagrams  $\{s, q\}$ and
$\{r, s\}$. From $\{s, q\}$ graph it can be seen that the best-fitted trajectory is currently between SCDM and SS state with $\{s, q\} = {1.065, -0.8}$. Also from $\{r, s\}$ graph we see that the universe passed the LCDM.\\
\begin{tabular*}{2.5 cm}{cc}
\includegraphics[scale=.35]{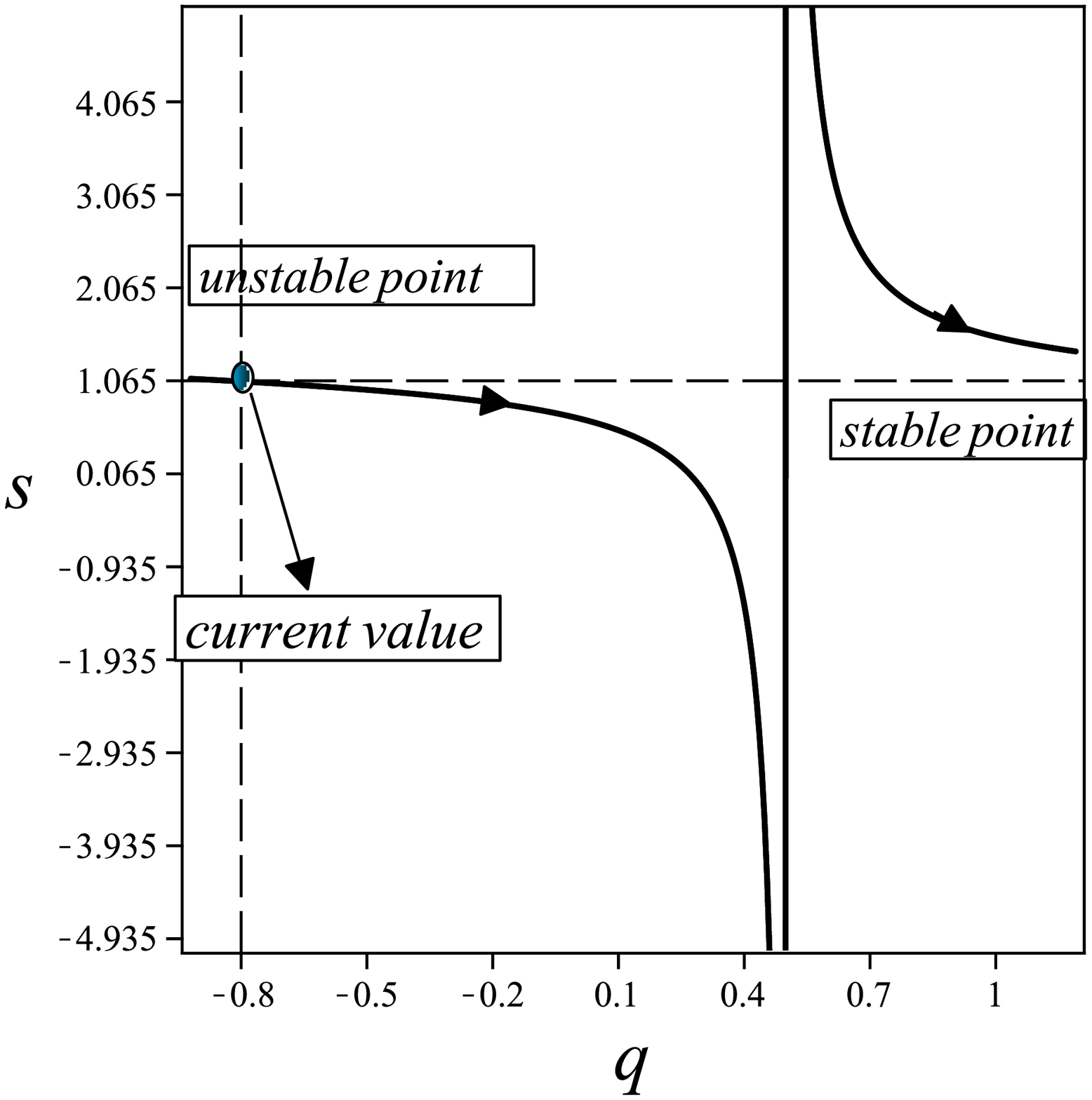}\hspace{0.1 cm} \includegraphics[scale=.35]{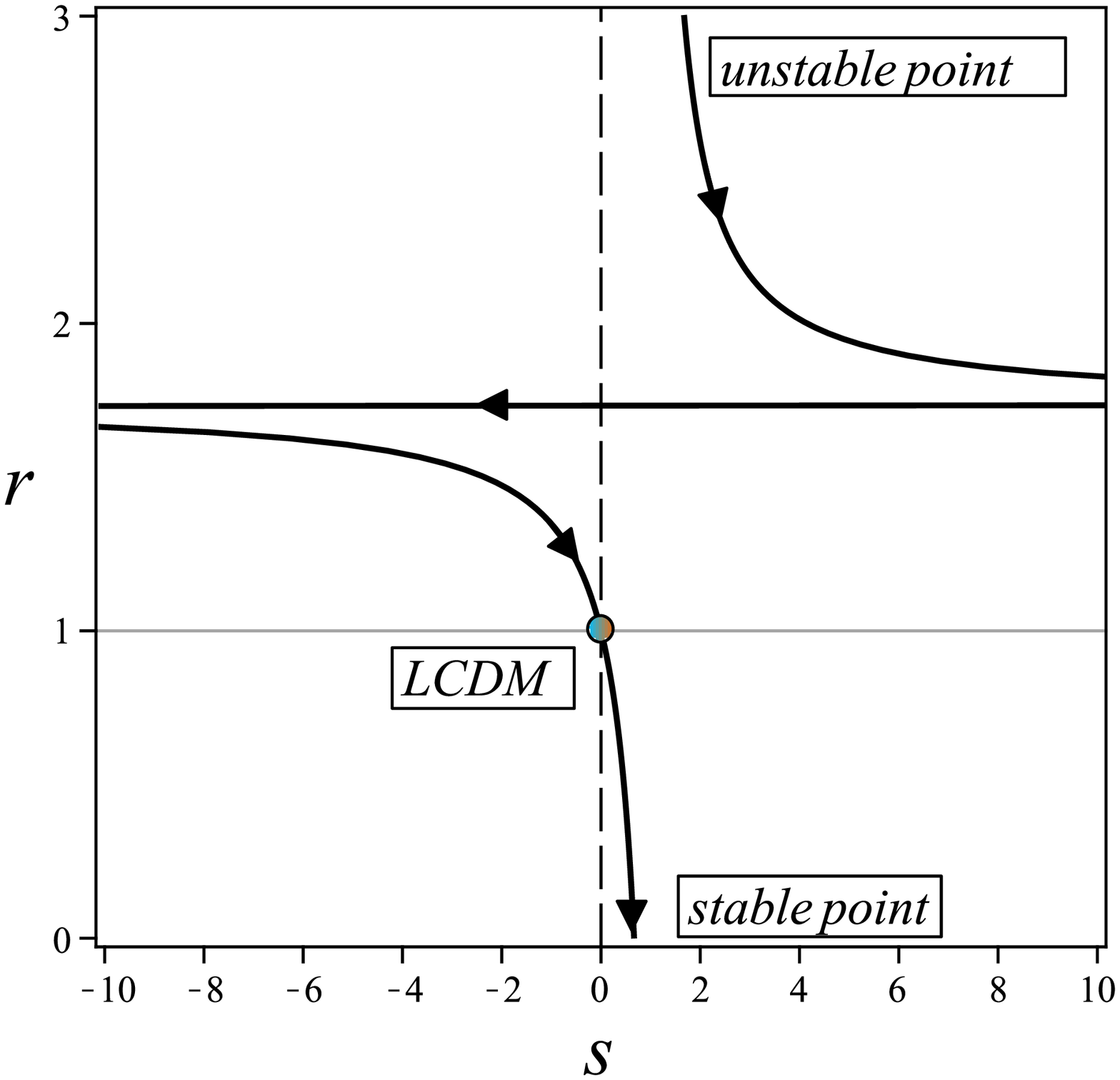}\hspace{0.1 cm} \\
Fig. 4:The best fitted statefinders parameters  $\{s- q\}$ and $\{r- s\}$\\.
\end{tabular*}\\

In Fig 5 we depict the corresponding dynamical behavior of the statefinder $\{r, s\}$ against
$N = ln(a)$. From Figs. 4 and 5 we observe that the universe starts its journey from unstable
state in the past, passed the current state and eventually reaches a stable state in the future.
Note that in Fig 4, the extreme points
of the statefinder trajectories corresponds to the state that the universe crosses phantom
divide line in future.\\

\begin{tabular*}{2.5 cm}{cc}
\includegraphics[scale=.35]{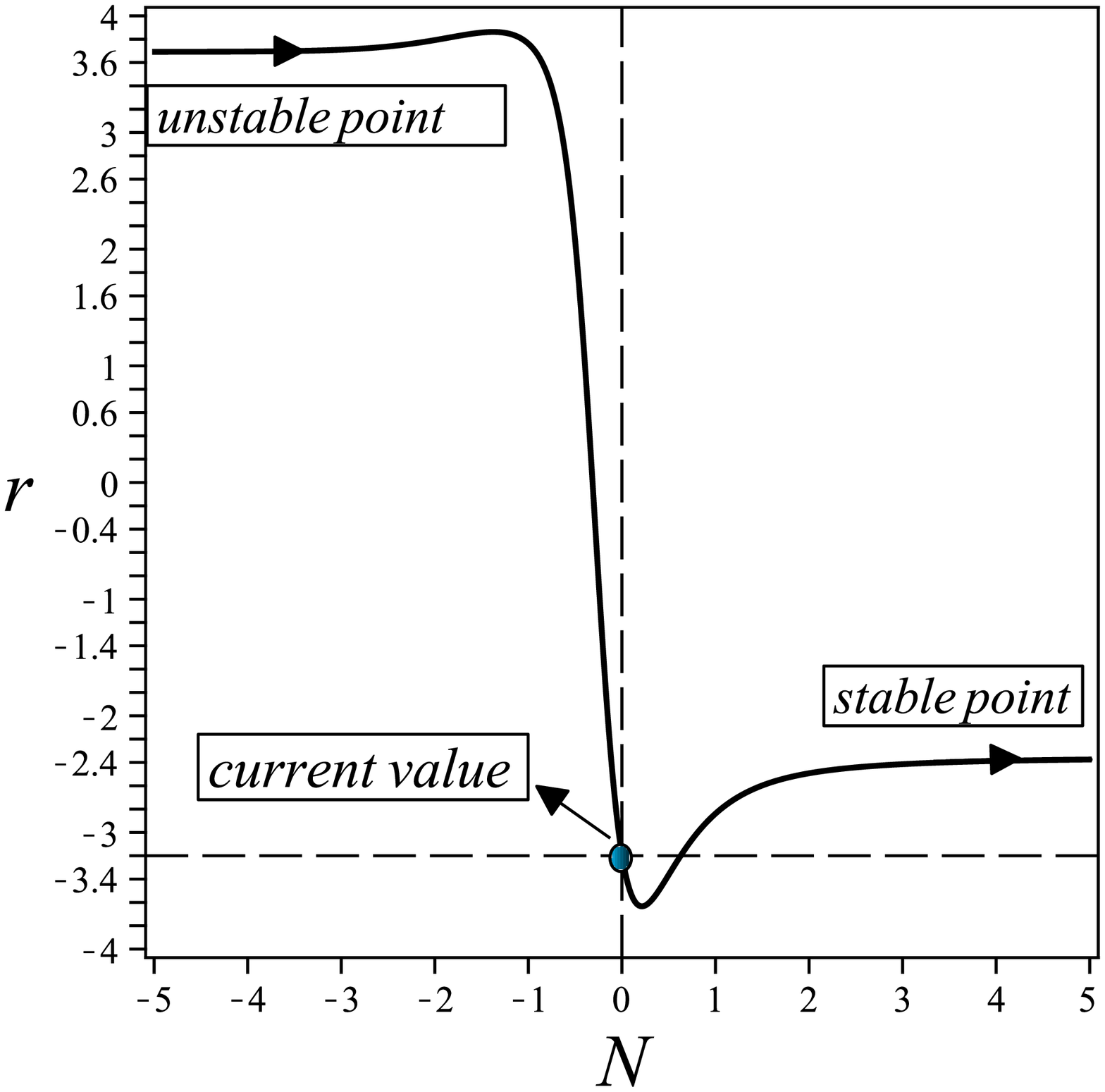}\hspace{0.1 cm}\includegraphics[scale=.35]{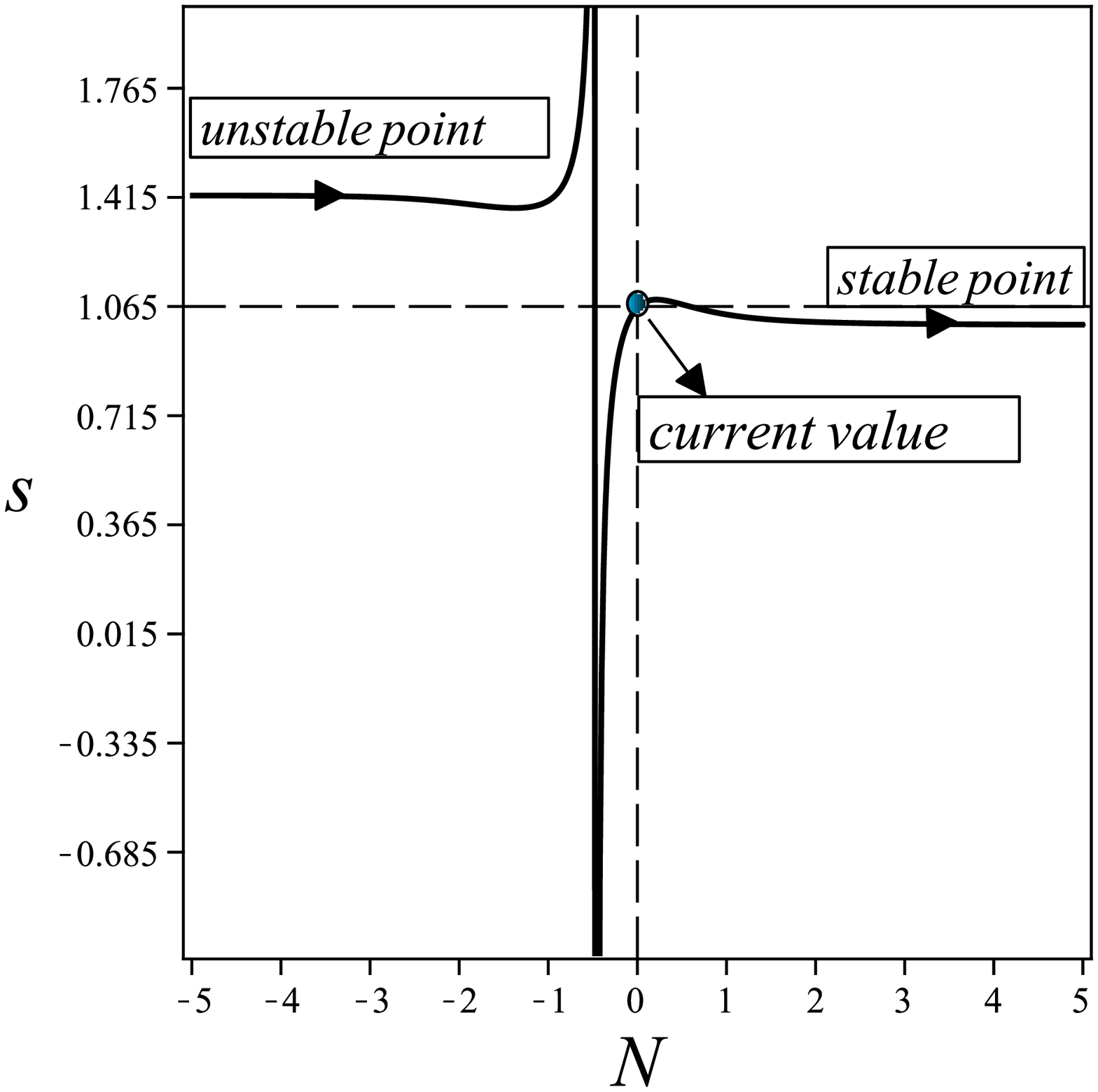}\hspace{0.1 cm}\\
Fig. 5:The best fitted statefinder parameters $r$ and $s$  plotted as function of $N=ln(a)$.\\
\end{tabular*}\\
fig 6 shows the dynamical behavior of the $\Omega_{\Lambda}$,$\Omega_{m}$,$\Omega_{b}$ and $\Omega_{k}$   for best fitted  both parameters and initial conditions\\
\begin{tabular*}{2.5 cm}{cc}
\includegraphics[scale=.4]{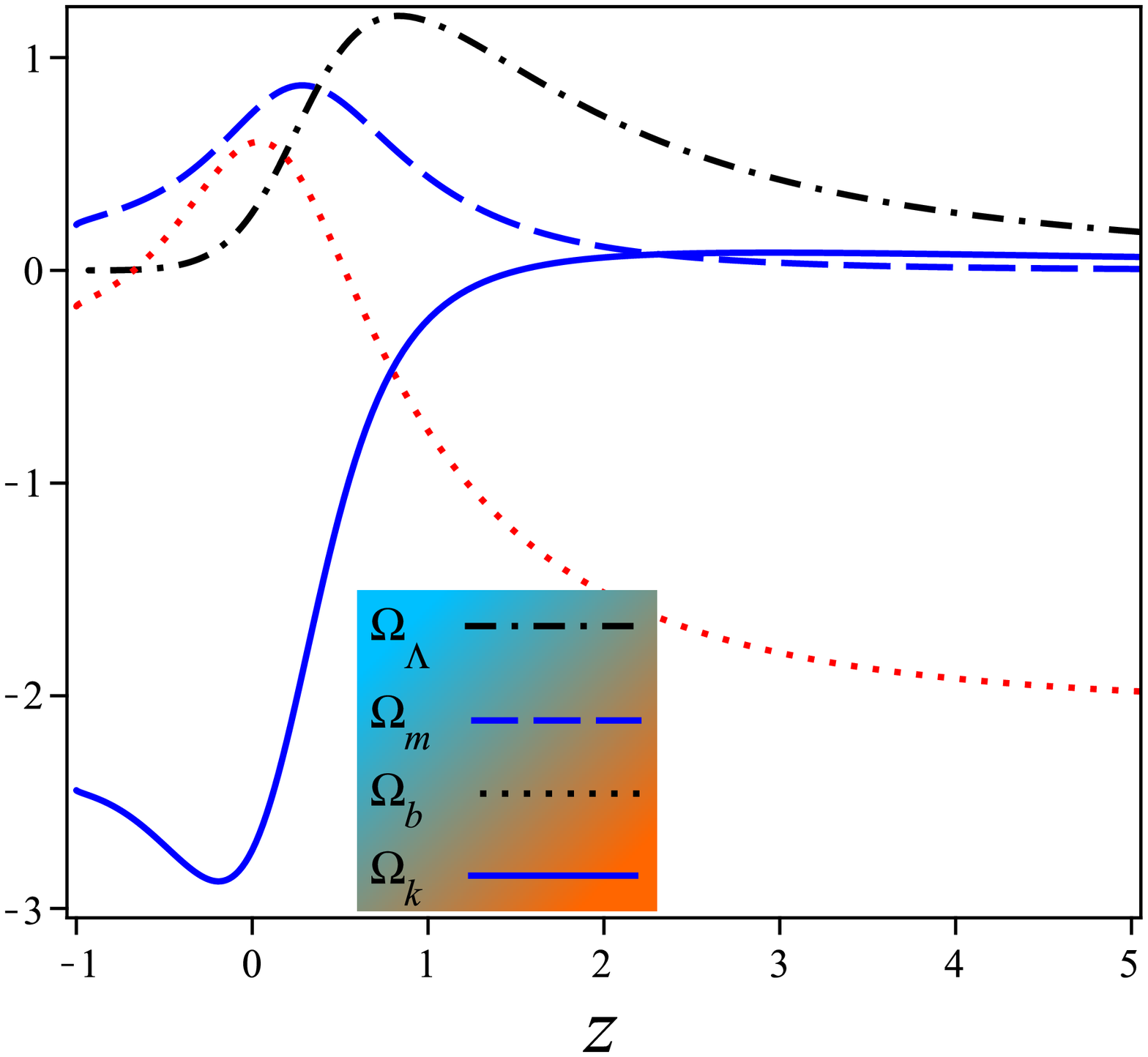}\hspace{0.1 cm}\\
Fig. 6: dynamical behavior of $\Omega_{\Lambda}$,$\Omega_{m}$,$\Omega_{b}$ and $\Omega_{k}$   plotted as functions of redshift  \\
\end{tabular*}\\\\
Fig.8 shows the dynamic of the  $(\omega_{eff})$ against redshift $z$ in the presence and absence of the extra dimension. As can be seen from Fig. 7, there is a slight difference between the trajectories of these two cases, we ca see from fig 7 that the current value of EoS parameter  in absence of extra dimension  is $(\omega_{eff})\simeq-0.65$ while  in presence of extra dimension $(\omega^{current}_{eff})\simeq-0.85$ which is more consistent with observational data\\
\begin{tabular*}{2.5 cm}{cc}
\includegraphics[scale=.4]{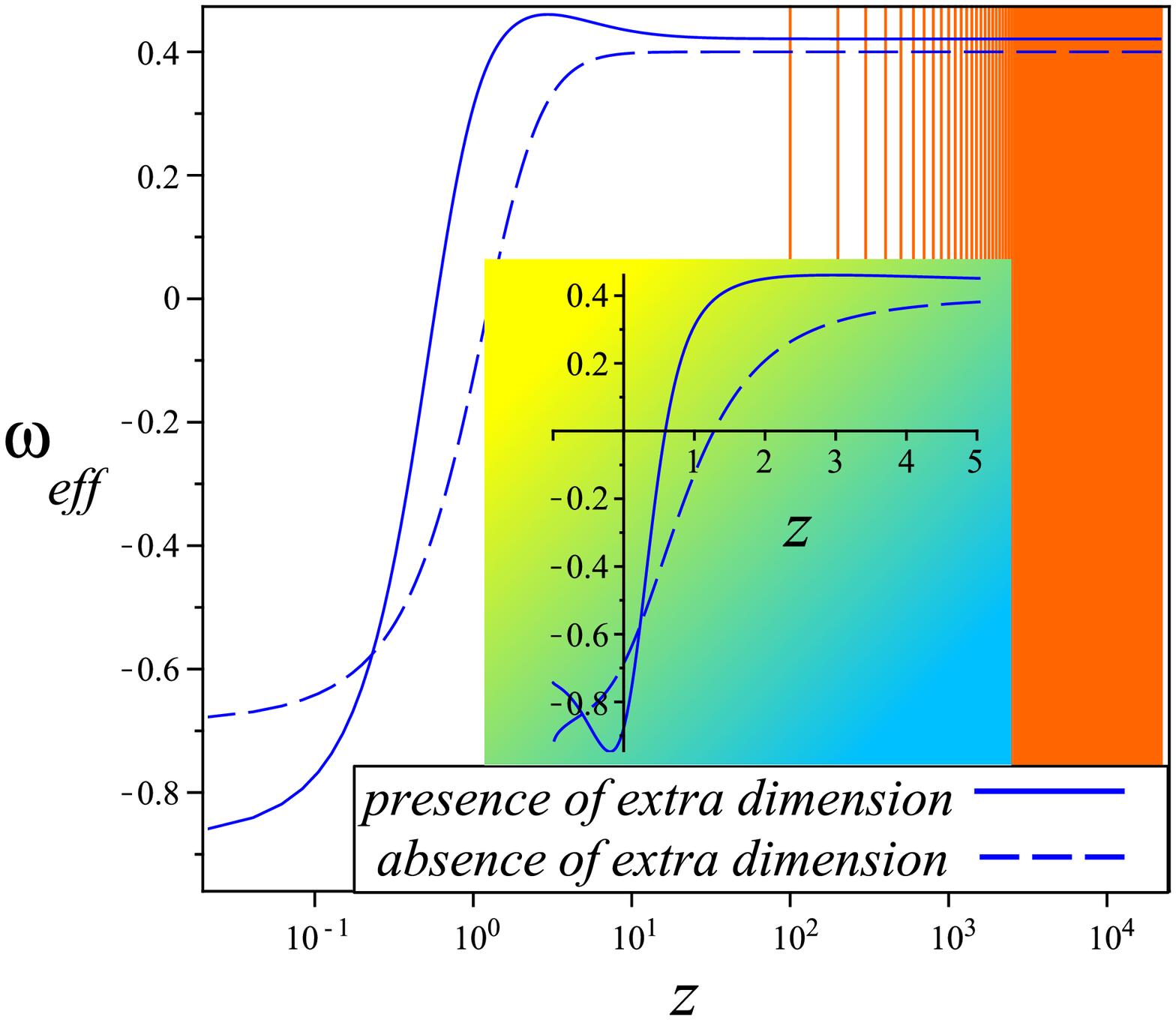}\hspace{0.1 cm}\\
Fig.8: dynamical behavior of effective EoS parameter  $(\omega_{eff})$  plotted as functions of redshift \\in the  solid line): presence of extra dimension and  dash line): absence of extra dimension\\
\end{tabular*}\\

Fig.9 shows The distance modulus $\mu(z)$ plotted as function of redshift for the best fitted parameters \\

\begin{tabular*}{2.5 cm}{cc}
\includegraphics[scale=.4]{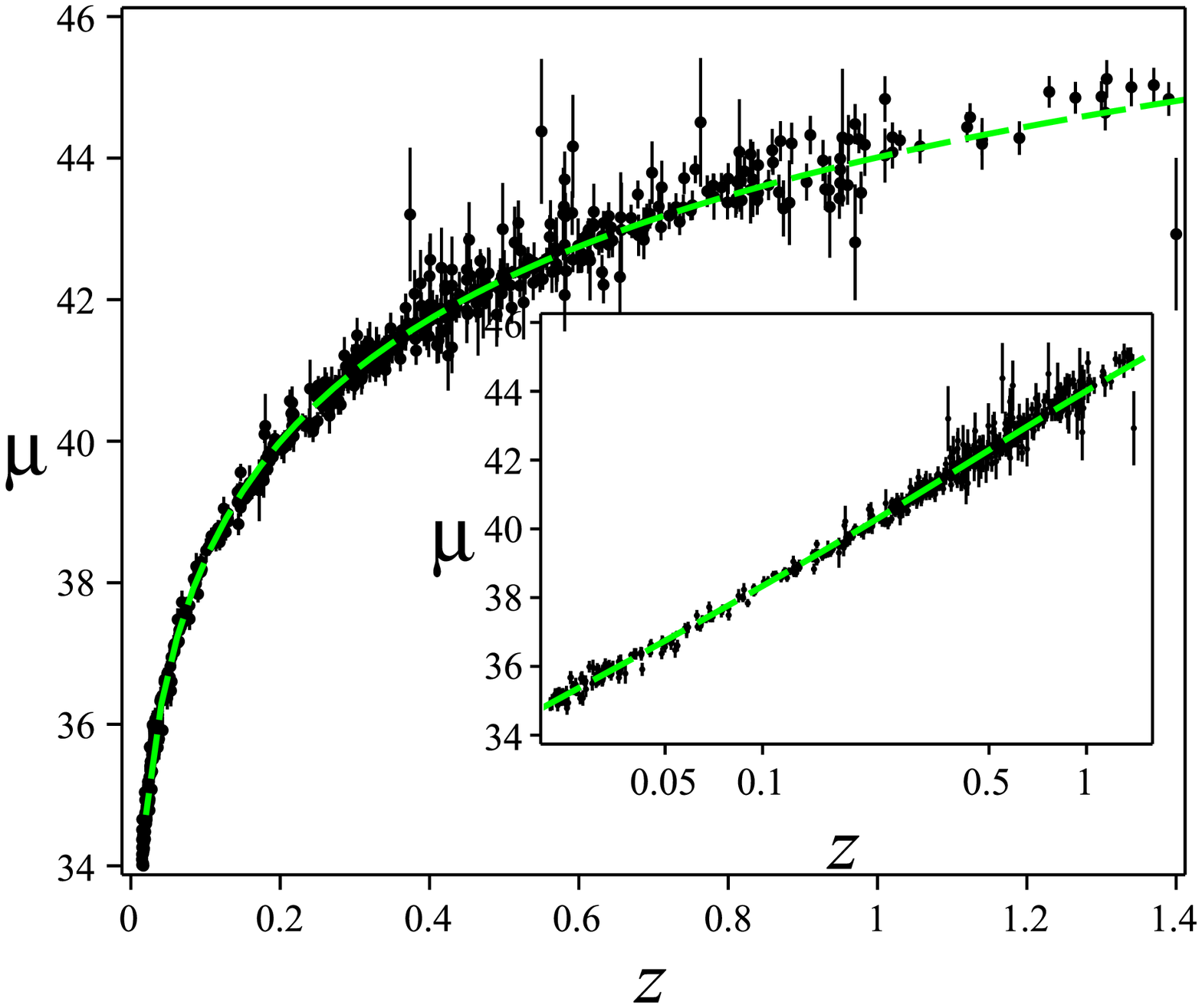}\hspace{0.1 cm} \\
The distance modulus $\mu(z)$ plotted as function of redshift for the best fitted parameters\\
\end{tabular*}\\
\section{Summary and discussion\label{CONC}}

We have constructed an autonomous phase-plane in 5D Brans-Dicke cosmology in the presence of agegraphic dark energy and cold dark matter. We have improved the phase-plane analysis by constraining the stability and model parameters by SNe Ia observational data for distance modulus using $\chi^2$ method. In a dark energy dominated universe and with the phase space and stability analysis, we find two critical points, and the best fitted trajectory from unstable critical point to the attractor. A quantitative analysis of the dynamical variables and physical parameters in the theory is presented. While both effective and ADE EoS parameters depend on the BD scalar field, model parameter $n$ and the extra dimension $b$, the best fitted trajectories are shown in Fig. 7 in the presence and absence of extra dimension. we see that in presence of extra dimension the dynamical evolution   of $\omega_{eff}$  is more consistent with observational data. The results show that the universe has passed through SCDM and LSDM phase and approaching the SS phase in future.



\begin{thebibliography}{}

\bibitem{Dunkley} Dunkley J. et al., (2009), Astrophys. J. 701, 1804;
 Gold B. et al., (2009), Astrophys. J. Suppl. 180, 265; Hill R.S. et al., (2009), Astrophys. J.
Suppl. 180, 246; Hinshaw G. et al., (2009), Astrophys. J. Suppl. 180, 225 ;  Nolta M.R.
et al, (2009), Astrophys. J. Suppl. 180, 296.

\bibitem{Komatsu} Komatsu E. et al., (2009), Astrophys. J. Suppl. 180, 330.

\bibitem{Knop} Knop R.A. et al., (2003), Astrophys. J. 598 102 ; Garnavish P.M. et al., (1998), Astrophys. J. 493,L53; Perlmutter S. et al., (1997),  Astrophys. J. 483,565; Frieman J.A. et al., (2008), Astrophys. J. 135,338 ;  Sako M. et al., (2008), Astrophys. J. 135,348.

\bibitem{Riess1} Riess A.G. et al., (2004), Astrophys. J. 607,665.

\bibitem{Leauthand} Leauthand A. et al., (2010), Astrophys. J. 709,97; Kubo J.M. et al., (2009), Astrophys. J. 702, L110;  Sato M. et al., (2009), Astrophys. J. 701,945

\bibitem{Parkinson} Parkinson D. et al., (2010), Mon. Not. Roy. Astron. Soc. 401,2169;  Percival W. et al., (2010), Mon. Not. Roy. Astron. Soc. 401,2148; Wang X. et al., (2009), Mon. Not. Roy. Astron. Soc. 394,1775;  Benitez  N. et al., (2009), Astrophys. J. 691,241.

\bibitem{Cole} Cole S. et al., (2005), Mon. Not. Roy. Astron. Soc. 362, 505; Croton D.J. et al., (2005), Mon. Not.
Roy. Astron. Soc. 356, 1155; Wild V. et al., (2005), Mon. Not. Roy. Astron. Soc. 356,247.

\bibitem{Yan} Yan, R. et al., (2009), Mon. Not. Roy. Astron. Soc. 398,735; Sawicki M. et al., (2008), Astrophys. J. Astrophys. J. 687,884; Schiavon R.P. et al., (2006), Astrophys. J. 651,L93.

\bibitem{Riess} Riess A.G. et al., (2004), Astrophys. J. 607,665.

\bibitem{Amanullah} Amanullah R. et al., arXiv:1004.1711v1, to appear in (2010), Astrophys. J.

\bibitem{Cai} Cai Y.-F. et al., (2010), Phys. Rept., in press, doi:10.1016/j.physrep.2010.04.001.

\bibitem{Gonzalez} Gonzalez T., Leon G, and Quiros I., (2006),  Classical Quantum Gravity 23,165; Johri V.B., (2002),  Classical Quantum Gravity 19,5959; Brax P. and Martin J., (2000), Phys. Rev. D 61,103502.

\bibitem{Carroll} Carroll S.M., De Felice A., and Trodden M., (2005), Phys. Rev. D 71,023525; Stefancic H., (2004)
Phys. Lett. B 586,5.

\bibitem{Caldwell} Caldwell R.R., Kamionkowski M., and Weinberg N.N., (2003), Phys. Rev. Lett. 91, 071301;
Caldwell R.R., (2002), Phys. Lett. B 545,23.

\bibitem{Cohen} Cohen A. G., Kaplan D. B. and Nelson A. E., (1999), Phys. Rev. Lett. 82,4971.

\bibitem{Hsu} Hsu S. D. H., (2004), Phys. Lett. B 594,13.

\bibitem{Li} Li M., (2004), Phys. Lett. B 603,1.

\bibitem{Hooft} Hooft G. 't , gr-qc/9310026 ; Susskind L., (1995), J. Math. Phys, 36,6377.

\bibitem{Cai1} Cai Y.-F. and Wang J., (2008), Class. Quant. Grav. 25,165014; Guo Z.-K. et al.,  (2005), Phys. Lett.
B 608,177; Feng B., Wang X.L., and Zhang X.M., (2005), Phys. Lett. B 607,35.

\bibitem{Xia} Xia J.Q. et al., (2009), Int. J. Mod. Phys. D 17, 1229 ; Kunz M. and Sapone D.,  (2006), Phys. Rev. D74, 123503; Caldwell R.R. and Doran M., (2005), Phys. Rev. D. 71,23515.

\bibitem{Copeland} Copeland E.J., Sami M., and Tsujikawa S., (2006), Int. J. Mod. Phys. D 15, 1753; Peebles
P.J.E. and Ratra B., (2003), Rev. Mod. Phys. 75,559.

\bibitem{Sahoo} Sahoo B.K. and Singh L.P., (2002), Mod. Phys. Lett. A 17,2409.

\bibitem{Boisseau} Boisseau B. et al., (2000), Phys. Rev. Lett. 85,2236.

\bibitem{Gannouji} Gannouji R. et al., (2006), J. Cosmol. Astropart. Phys. 0609,016.

\bibitem{Sahoo1} Sahoo B.K. and Singh L.P., (2003), Mod. Phys. Lett. A 18, 2725.

\bibitem{Sahoo2} Sahoo B.K. and Singh L.P., (2004), Mod. Phys. Lett. A 19, 1745.

\bibitem{Sadeghi} Sadeghi J., Setare M.R., Banijamali A. and Milani F., (2009), Phys. Rev. D 79, 123003.

\bibitem{Capozziello} Capozziello S., Carloni S. and Troisi A., (2003), Recent Res. Dev. Astron. Astrophys. 1 625.

\bibitem{Nojiri} Nojiri S.  and  Odintsov S.D., (2003), Phys. Rev. D 68  123512; (2003), Phys. Lett. B 576, 5; Farajollahi H., Salehi A., Tayebi F., Ravanpak A., (2011), J. Cosmol. Astropart. Phys. 05, 017.

\bibitem{Faraoni} Faraoni V., (2007), Phys. Rev. D 75  067302; Briscese F., Elizalde E., Nojiri S. and Odintsov S.D., (2007), Phys. Lett. B646, 105.

\bibitem{Nojiri3} Nojiri S. and Odintsov S.D., (2004), Gen. Rel. Grav. 36, 1765; (2004), Phys. Lett. B 599, 137; Farajollahi H., Milani F., (2010), Mod. Phys. Lett. A 25:2349-2362; Farajollahi H., Setare M. R., Milani F. and Tayebi F., (2011), Gen.Rel.Grav.43:1657-1669.

\bibitem{Setare1} Setare M. R., Jamil M., (2010), Phys. Lett. B 690  1-4;
Ito Y., Nojir S. i, (2009), Phys.Rev.D79:103008.

\bibitem{Mota1} Mota D.F., Shaw D.J., (2007), Phys. Rev. D 75, 063501.

\bibitem{Dimopoulos} Dimopoulos K., Axenides M., (2005), J. Cosmol. Astropart. Phys. 0506:008.

\bibitem{Damouri} Damour T., Gibbons G. W. and Gundlach C., (1990), Phys. Rev. Lett, 64, 123; Farajollahi H., Mohamadi N., (2010), Int. J. Theor. Phys.49:72-78; Farajollahi H., Mohamadi N. , Amiri H., (2010), Mod. Phys. Let. A, Vol. 25, No. 30 2579–2589.

\bibitem{Carr} Carroll S. M., (1998), Phys. Rev. Lett. 81 3067.

\bibitem{carrolll} Carroll S. M., Press W. H. and Turner E. L., (1992), Ann. Rev. Astron. Astrophys, 30, 499.

\bibitem{Biswas} Biswas T. and Mazumdar A., arXiv:hep-th/0408026.

\bibitem{Biswass} Biswas T., Brandenberger R., Mazumdar A. and Multamaki T., (2006), Phys.Rev. D74 , 063501.

\bibitem{Sahni} Sahni V., Saini T. D., Starobinsky A. A., Alam U., (2003), JETPLett.77:201-206.

\bibitem{Linder} Linder E.V., (2008), Rep. Prog. Phys. 71,  056901.

\bibitem{Khourym} Khoury J. and Weltman A., (2004), Phys. Rev. Lett. 93, 171104.

\bibitem{Brax2} Brax P., de Bruck C van , Davis A. C., Khoury J., Weltman A., (2004), Phys.Rev.D70:123518 ; Khoury J., Weltman A., (2004), Phys. Rev. D 69, 044026.

\bibitem{farajollahi} Farajollahi H., Salehi A., (2010), Int. J. Mod. Phy. D, 19(5), 1–13.

\bibitem{Brans} Brans C. H. and Dicke R. H., (1961), Phys. Rev. 124, 925.

\bibitem{mtheory7} Bouhmadi-Lopez M., (2008), Nucl. Phys. B. 797, 78-92.

\bibitem{mtheory} Agarwal N., Bean R., Khoury J. and Trodden M., (2010), Phys. Rev. D. 81, 084020.

\bibitem{mtheory3} Lee H. M. and Tasinato G., (2004), J. Cosmol. Astropart. Phys. 0404, 009.

\bibitem{mtheory4} Karasik D. and Davidson A., (2004), Class. Quant. Grav. 21, 1295-1302.

\bibitem{mtheory5} Aoyanagi K. and Maeda K., (2006), J. Cosmol. Astropart. Phys. 0603, 012.

\bibitem{mtheory6}Shtanov Y., Viznyuk A. and Sahni V., (2007), Class. Quant. Grav. 24, 6159-6190.

\bibitem{Q} Huang Q. G. and  Li M., (2004), J. Cosmol. Astropart. Phys. 0408, 013;(2005), J. Cosmol. Astropart. Phys. 0503, 001; Setare M. R., Zhang J. and  Zhang X., (2007), J. Cosmol. Astropart. Phys. 0703,007.

\bibitem{Wang} Wang  B., Gong Y. G. and  Abdalla E., (2005), Phys. Lett. B 624, 141; Wang B., Lin C. Y.
and  Abdalla E., (2006), Phys. Lett. B 637,357.

\bibitem{X} Zhang X. and Wu F. Q., (2005), Phys. Rev. D 72, 043524; (2007), Phys. Rev. D 76, 023502;
 Chang Z., Wu F. Q., and  Zhang X., (2006), Phys. Lett. B 633,14.

 \bibitem{R} Cai R. G., (2007), Phys. Lett. B 657,228 .

\bibitem{Sahni} Sahni V., Saini T. D., Starobinsky A. A.,  Alam U., (2003), JETPLett.77:201-206.

\bibitem{Alam}  Alam U., Sahni  V.,  Saini T. D. and  Starobinsky A. A.,(2003), Mon. Not. Roy. Astron. Soc. 344, 1057.
\bibitem{Zimdahl} Zimdahl W.,  Pavon D., (2004), Gen. Rel. Grav. 36, 1483.

\bibitem{Yi}  Yi Z. L. and  Zhang T. J., (2007), Phys. Rev. D75, 083515.

\bibitem{H. Farajollahi}  Farajollahi H., Salehi  A., (2011), Phys.Rev.D83:124042;  Farajollahi H., Salehi A., Tayebi F., Ravanpak A., (2011), JCAP 05 017;  Farajollahi Hossein,  Salehi Amin, (2010), JCAP 1011:006;
 Farajollahi H., Salehi  A., Tayebi F., Astrophys Space Sci (2011) 335:629-634

\bibitem{Qiang} Qiang, L.Y.Ma, and Yu D., (2005), phys. Rev. D 71,061501.

\bibitem{Over} Overdium J. M. and Wesson P. s., (1997), phys. Rep. 283,303.



\end{thebibliography}
\end{document}